[*Note added for astro-ph:* In this short introductory commentary on the paper (Hessels et al 2006, Science, 311, 1901) reporting the discovery of the shortest spin period millisecond pulsar (MSP) Ter5-ad in the globular cluster Terzan 5, I also point out a new explanation for possible minimum spin periods, *P,* of MSPs without requiring gravitational radiation (or other) slow-down torques. If the accretion of matter required to spinup a MSP also reduces (buries) the neutron star (NS) magnetic field, *B*, as commonly believed, an inverse correlation between neutron star mass, *M*, and *B* is expected together with a positive correlation between *P* and *B*. Both are suggested for the 4 MSPs with NS mass measures reported (Latimer and Prakash 2004, Science, 304, 536) to have ≤10% uncertainties. The correlations imply the Ter5-ad NS has ~2.5 $M_\odot$, $B$ ~5 x $10^7$ G and thus $\dot{P}$ ~3 x $10^{-21}$ s/s – which can be tested when a timing solution is found. If confirmed, the highest spin frequency NSs do not pulse simply because their *B* fields are too low.]

# A Neutron Star in F-sharp

Jonathan E. Grindlay[1]

Millisecond pulsars are extreme examples of what can happen when stars evolve into neutron stars in compact binary systems. These rotating objects are spun up by accretion of matter from their binary companions, producing luminous X-ray emission, and later become detectable as pulsars with periods of a few milliseconds (1). As a result, these "fast pulsars" may offer some of the best probes to study matter and space in the relativistic regime of strong gravity. On page 1901, Hessels et al (2) report the discovery of pulsar PSR J1748-2446ad in the dense globular cluster Terzan 5 (Ter5-ad). This object, detected with the Green Bank radio Telescope, holds the new record for the fastest spinning neutron star (or indeed any object of stellar mass or larger). Its spin period is only 1.396 ms, even shorter than that of B1937+21 [the first millisecond pulsar discovered (3)] at 1.558 ms. With a rotation frequency of 716 Hz, Ter5-ad reaches a new high note for the music of the celestial spheres – between F and F sharp, whereas B1937+21 (at 642Hz) can only hit a note between D-sharp and E.

Since their discovery in 1967, pulsars have been the gateway to the study of matter and energy at the extremes found only in neutron stars (4). Such stars are nature's last stable outposts of matter and only a factor of ~3 larger in radius than what would collapse to a black hole. With ~1.4-2 solar masses ($M_\odot$) packed into a ~10-15 km radius, neutron stars are the ultimate laboratories for astrophysics and physics of the extreme. Neutron stars can exhibit magnetic fields about $10^{15}$ times that of the Earth, as revealed in giant flares of magnetars. And neutron star – binary pairs merge to produce extremely energetic events as revealed in short gamma-ray bursts. However it is the oldest, and fastest pulsars, the millisecond pulsars, that may allow the most direct measures of the ultimate prize: the mass $M$ and radius $R$ of the neutron star itself, which would fix the equation of state and the composition of matter at hypernuclear density. The Ter5-ad system is a new stepping stone to the quest for $M$ and $R$ as well as a constraint on the ultimate rotational limits that may be revealed by gravitational waves.

[1] The author is at the Harvard-Smithsonian Center for Astrophysics, 60 Garden St., Cambridge, MA 02138. E-mail: josh@cfa.harvard.edu

A point on the rotation equator of Ter5-ad has velocity nearly one-fourth the speed of light, assuming $R = 15$ km. The constraint that the maximum spin frequency $\nu_s$ not exceed the Keplerian orbital frequency $\nu_K$ at the neutron star surface gives the simple constraint that $\nu_s \leq 1833(M/M_\odot) R_{10}^{-3/2}$ Hz where $M$ is the neutron star mass and $R_{10}$ is the radius in units of 10 km. Taking into account general relativistic effects, Lattimer and Prakash (5) derived a value of 1045 Hz as the maximum spin frequency for a neutron star with (non-rotating) radius $R$ and mass $M$ less than the maximum mass allowed by its equation of state. A measured spin frequency $\nu_s$ then sets an upper bound on the neutron star radius. For Ter5-ad with $\nu_s = 716$Hz, $R$ is restricted to be $\leq 14.4$ to 16.7 km when $M$ is $\leq 1.4$ to 2.2 $M_\odot$, which is the approximate range encompassed by recent neutron star mass measurements with quoted uncertainties $\leq$10% (5). As Hessels et al point out, a mass measurement -- and thus a constraint on the equation of state -- for the neutron star in Ter5-ad is conceivable given its eclipsing geometry if the radial velocity of the $\geq 0.14$ $M_\odot$ main-sequence binary companion can be measured. Given likely stellar crowding, this will be difficult, however: With the ~7 magnitudes of optical extinction to Terzan 5, the companion must be sought in the in the near-infrared with an expected infrared magnitude as faint as ~25.

X-ray observations allow complementary constraints on M and R for the neutron stars in quiescent low mass X-ray binaries (qLMXBs) as well as their millisecond pulsar descendents. The first Chandra X-ray survey of the globular cluster 47Tuc revealed that source X7 was one of several qLMXBs (6). Analysis of deeper Chandra observations of X7 with fits of neutron star atmosphere models (including surface gravities) to its purely thermal X-ray emission spectrum can be made. These fits yield 90% confidence limits ranging from $R = 12.7$ to 16.7 km (for $M = 1.4\ M_\odot$) to $R = 10.0$ to 15.0 km (for $M = 2.2\ M_\odot$) (7), which are consistent with those for Ter5-ad. Even more accurate constraints of the gravitational redshift factor M/R (which in turn constrains the equation of state given measures of $M$) for neutron stars are possible from studies of the pulsed profiles of thermal X-ray emission from millisecond pulsars, provided the pulsar distance and inclination angles are known (8). Unfortunately Ter5-ad may not allow this, since its X-ray emission is likely dominated by unpulsed non-thermal emission due to the shock where its pulsar wind encounters gas from its main sequence binary companion, as recently identified (9) in the similar millisecond pulsar 47Tuc-W.

Hessels et al suggest that even faster millisecond pulsars exist but may be hidden by the increased likelihood of radio eclipses, because their increased spin-down energy loss rate and resulting pulsar wind more readily drives matter off their binary companions. Indeed the distribution (see the figure) of the energy flux from the spin-down energy loss incident on the binary companion for the 12 fastest millisecond pulsars with binary companions (10) shows that the eclipsing (radio) systems are generally those with the largest incident flux. Still faster millisecond pulsars are preferentially hidden (if still in binaries) from radio surveys but would be detectable as relatively hard X-ray sources (unpulsed) from their pulsar wind shocks.

A maximum NS spin frequency, $v_{max} \sim 760$ Hz, was suggested (11) given the spin distribution for accretion-powered as well as radio millisecond pulsars. A value for $v_{max}$ much less than 1045Hz, the maximum allowed independent of the equation of state (5), could require a source of angular momentum loss such as gravitational radiation from an r-mode instability in the neutron star core (12). Another mechanism may be at work: if the accreting matter spinning up the neutron star is burying its primordial magnetic field, -- as is generally believed (1) to account for the $B \leq 10^9$ G fields inferred for qLMXBs and millisecond pulsars -- then the increased accretion and final spin $v_s$ imply a lower $B$ field emerging from the neutron star when accretion stops. The millisecond pulsars that spin fastest would then have the lowest $B$ fields (above threshold for pulsations) and largest neutron star mass; neutron stars with larger values of $v_s$ and mass have smaller $B$ and do not pulse. For those millisecond pulsars with neutron star mass estimates (5), including the $2.2 \pm 0.2\ M_\odot$ value for the neutron star in the millisecond pulsar J0751+1807 (13), possible correlations between $B$ vs. $P$ and $B$ vs. $M$ are evident (see the figure). The implied values for Ter5-ad are $\sim 2.5\ M_\odot$ and $\sim 7 \times 10^7$ gauss and thus a predicted change in spin period $\dot{P} \sim 3 \times 10^{-21}$ s/s, which can be tested when a final timing solution is found. If this were the case, then Ter5-ad would have a mass approaching the maximum (5) for neutron stars, and it could be singing nearly the highest note without having made gravitational waves.

**References**
1. D. Bhattacharya and E.P.J. van den Heuvel *Phys. Rep.* **203,** 1 (1991).
2. J.W.T. Hessels *et al., Science* **311,** xxx (2006); published online 12 January 2006 (10.1126/science, 1123430)..
3. D.C. Backer, S.R. Kulkarni, C. Heiles, M.M. Davis, W.M. Goss, *Nature* **300,** 615 (1982).
4. Special issue on pulsars, *Science* **304** (23 April 2004).
5. J.M. Lattimer and M. Prakash, *Science* **304,** 536 (2004).
6. J.E. Grindlay, C.O. Heinke, P.D. Edmonds and S.S. Murray, *Science* **292,** 2290 (2001); published online 17 May 2001 (10.1126/science.1061135).
7. C.O. Heinke, G.B. Rybicki, R. Narayan and J.E. Grindlay, *Astrophys. J.,* in press (2006).
8. G.G. Pavlov and V.E. Zavlin, *Astrophys. J ,***490,** L91 (1997).
9. S. Bogdanov, J.E. Grindlay, and M. van den Berg, *Astrophys. J.* **630,** 1029 (2005).
10. Australia Telescope National Facility Pulsar Catalogue (www.atnf.csiro.au/research/pulsar/psrcat/).
11. D. Chakrabarty *et al., Nature* **424,** 42 (2003).
12. R. V. Wagoner, *Astrophys. J.* **278,** 345 (1984).
13. D.J. Nice *et al., Astrophys. J.* **634,** 1242 (2005).

**Figures:**

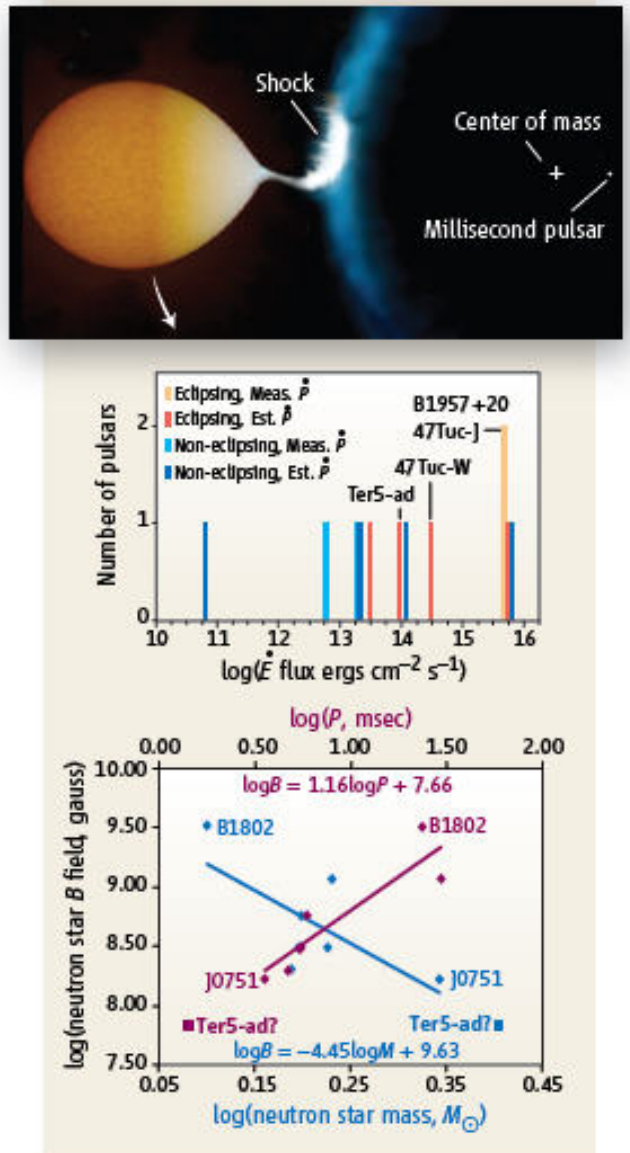

**Pulsar properties: (Top)** Gas from the normal star in the binary system is prevented from accreting onto the neutron star at the shock formed where it meets the "wind" of relativistic particles. The shocked gas eclipses the pulsar for much of the time so that pulsars with the shortest spins (strongest wind) and closest companions may be permanently hidden at radio frequencies, although unpulsed X-rays are expected. [Adapted from (9)] **(Center)** Pulsar wind energy flux of the 12 fastest spinning millisecond pulsars in binary systems that would be incident on their binary companion stars [parameters from (2) and (10)]. For pulsars without measured Pdots (including Ter5-ad), an estimated fixed value of $2 \times 10^{-20}$ s/s is used. The eclipsing systems generally do have higher pulsar wind flux values though Ter5-ad is not extreme, which implies that still faster systems could be found. The second-fastest millisecond pulsar, B1937+21, cannot be plotted because it has no binary companion. The other pulsars marked are B1957+20 (Black Widow), 47Tuc-J, and 47Tuc-W, which are numbers 3, 10, and 12 in order of increasing spin period (10). The two highest pulsar wind flux pulsars are the eclipsing system Ter5-O and the noneclipsing system M62-C. **(Bottom)** Correlations for millisecond pulsars with neutron star mass [values from (5)]. Correlation between mass and neutron star magnetic field $B$ [blue, from (10)], and between $B$ and spin period $P$ (purple). Extrapolated values are predicted for Ter5-ad; two other millisecond pulsars [see (10)] are marked for reference.